\documentclass[useAMS,referee,usenatbib]{biom}
\def\bSig\mathbf{\Sigma}

\usepackage[figuresright]{rotating}
\usepackage{graphicx}
\usepackage{amsmath}

\title[Latent ID ASCR with MCEM]{Approximate Maximum Likelihood Inference for Acoustic Spatial Capture-Recapture with Unknown Identities, Using Monte Carlo Expectation Maximization}

\author{Yuheng Wang$^{1,*}$\email{yw99@st-andrews.ac.uk}, 
Juan Ye$^{2,**}$\email{jy31@st-andrews.ac.uk}, Weiye Li$^{2,***}$\email{wl44@st-andrews.ac.uk}, and 
David L. Borchers$^{1,****}$\email{dlb@st-andrews.ac.uk} \\
$^{1}$Centre for Research into Ecological and Environmental Modelling,  School of Mathematics \\ and Statistics, University of St Andrews, The Observatory, St Andrews, Fife, KY16 9LZ, \\ Scotland. \\
$^{2}$School of Computer Science, University of St Andrews,  North Haugh, St Andrews, Fife, \\ KY16 9SX, Scotland.}

\begin{document}


\date{{\it Received October} 2023. {\it Revised February} 2023.  {\it
Accepted March} 2023.}



\pagerange{\pageref{firstpage}--\pageref{lastpage}} 
\volume{64}
\pubyear{2023}
\artmonth{November}


\doi{nothing!}


\label{firstpage}


\begin{abstract}
Acoustic spatial capture-recapture (ASCR) surveys with an array of synchronized acoustic detectors can be an effective way of estimating animal density or call density. However, constructing the capture histories required for ASCR analysis is challenging, as recognizing which detections at different detectors are of which calls is not a trivial task. Because calls from different distances take different times to arrive at detectors, the order in which calls are detected is not necessarily the same as the order in which they are made, and without knowing which detections are of the same call, we do not know how many different calls are detected. We propose a Monte Carlo expectation-maximization (MCEM) estimation method to resolve this unknown call identity problem. To implement the MCEM method in this context, we sample the latent variables from a complete-data likelihood model in the expectation step and use a semi-complete-data likelihood or conditional likelihood in the maximization step. We use a parametric bootstrap to obtain confidence intervals. When we apply our method to a survey of moss frogs, it gives an estimate within 15\% of the estimate obtained using data with call capture histories constructed by experts, and unlike this latter estimate, our confidence interval incorporates the uncertainty about call identities. Simulations show it to have a low bias (6\%) and coverage probabilities close to the nominal 95\% value.  \\

\end{abstract}

%



\maketitle


%

\section{Introduction}
\label{s:intro}

Passive acoustic monitoring (PAM) has become increasingly popular for surveying wildlife populations that are visually cryptic but acoustically detectable \citep{Gibb2019}. Detection of target species sounds in recordings using machine learning (ML) algorithms makes it possible to process large volumes of acoustic data quickly and robustly. One way of obtaining estimates of density from such acoustic data is to use acoustic spatial capture-recapture (ASCR) methods \citep{https://doi.org/10.1111/j.1365-2664.2009.01731.x, https://doi.org/10.1890/08-1735.1, stevenson2015}. These utilize the acoustic capture histories from the detector array (i.e., which calls are detected by which detectors) for density estimation. Constructing the capture histories from recordings (i.e., deciding which detections by different detectors are detections of the same call) can be challenging and time-consuming, and capture histories are essential for estimating density using ASCR methods. While automated methods exist for identifying vocalizations in acoustic recordings, and for incorporating uncertainty in vocalization identities into density estimation \citep*{wang2023automated}, no reliable automated method currently exists for constructing acoustic capture histories. In this paper, we develop a statistical inference algorithm for estimating the population density without known capture histories.

We consider three kinds of SCR likelihood functions in this paper: complete-data likelihoods, semi-complete-data likelihoods, and conditional likelihoods. Complete-data SCR likelihoods \citep[see, e.g.][]{https://doi.org/10.1890/07-0601.1} include the locations of detected and undetected individuals as latent variables. Semi-complete data likelihoods \citep{10.1214/15-AOAS890} include the locations of detected individuals as latent variables but integrate out the locations of undetected individuals. Both of these have the total number of animals, $N$ as a parameter. Conditional SCR likelihoods \citep{borchers2008} are conditioned on the number of individuals detected and do not have $N$ as a parameter. They cannot therefore on their own be used to estimate abundance $N$ (or density) and are usually used in conjunction with a Horvitz-Thompson estimator of $N$. See \citet{10.2307/24780847} for a fuller description of these three kinds of SCR likelihood.
Whichever likelihood is used, standard SCR inference requires a known capture history matrix, indicating which detections on different detectors are of the same call \citep[see][for a review of SCR]{10.2307/24780847}.

\begin{figure}[]   
         \centering
         \includegraphics[width=6.5in]{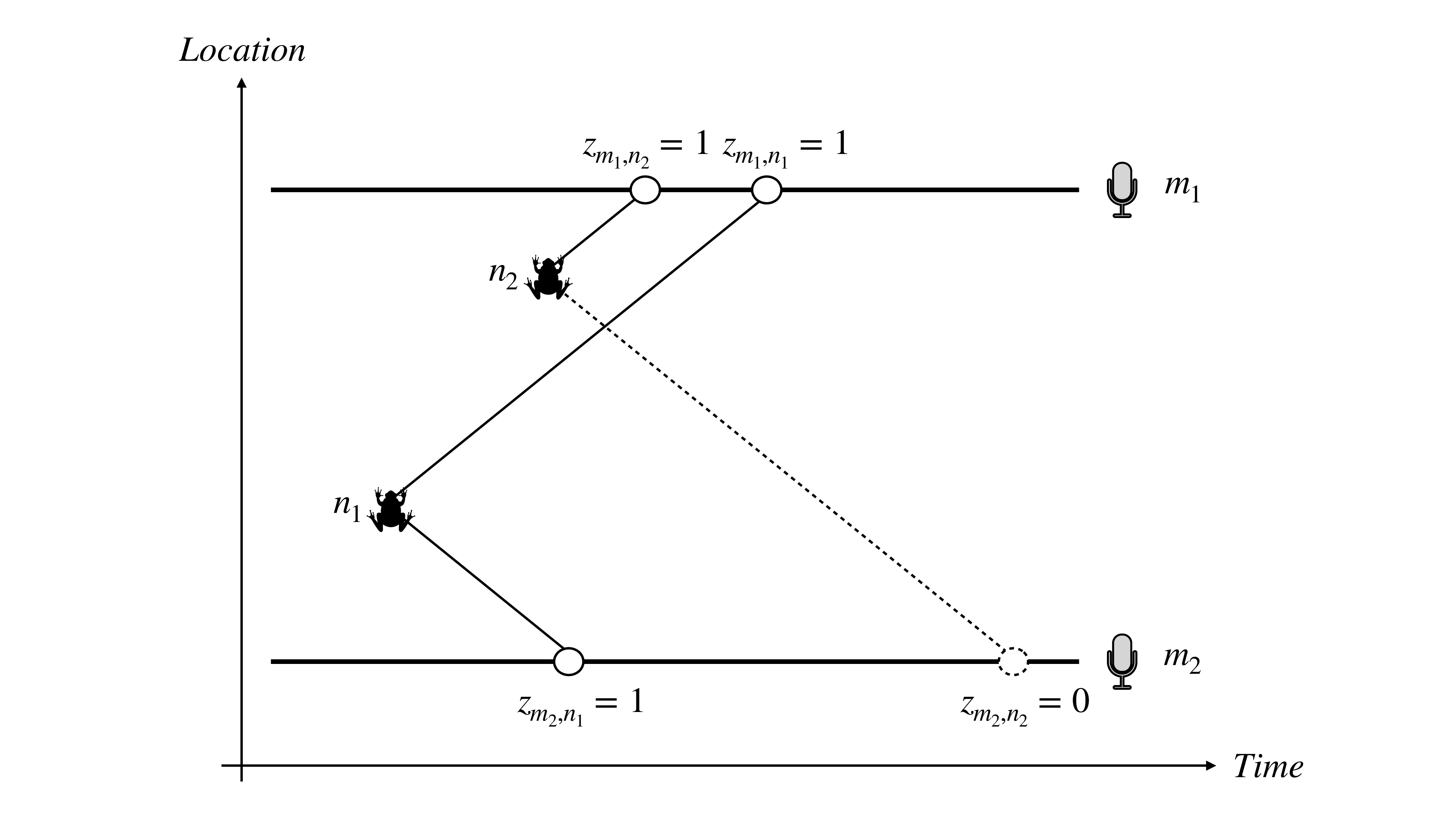}
          \caption{A demonstration of one-dimension space detection process. The $y-axis$ represents the one-dimensional location, and the $x-axis$ represents the time. The $m_1$ and $m_2$ are two detectors placed at different locations, corresponding to two recordings across time represented by two horizontal lines. The $n_1$ and $n_2$ are calls that are produced at two different locations and times, where $n_1$ comes first in time and closer to $m_2$, and $n_2$ comes later and closer to $m_1$ in distance. The slashes connecting the call source and the detector represent time consumption by sound propagation corresponding to the distance between the call and the detector, where the rate between distance and time consumption is a constant of sound velocity.   
          The detector $m_1$ detect both $n_1$ (i.e. $z_{m_1,n_1} = 1$) and $n_2$ (i.e. $z_{m_1,n_2} = 1$), while the $m_2$ only detect $n_1$ (i.e. $z_{m_2,n_1}=1$, $z_{m_2,n_2} = 0$). }
         \label{fig:align_demo}
\end{figure}

In what follows, we focus on the estimation of call abundance from an ASCR survey (the number of calls made over the duration of the survey) rather than animal density. See \citet{stevenson2015} for methods of converting call abundance to animal abundance or density, using supplementary data on mean call rate. On most acoustic SCR surveys, we do not really know the call capture histories. 
We usually know only the times (and possibly other features) of detections on each detector, not which detections are of the same call. We illustrate the problem in Fig \ref{fig:align_demo}. Suppose that we have two detectors $m_1$ and $m_2$ placed at different locations and two calls $n_1, n_2$ emitted at different locations, that are close together in time. Detector $m_1$ detects both $n_1$ and $n_2$ (indicated by $z_{m_1, n_1} = 1, z_{m_1, n_2} = 1$), while detector $m_2$ detects only call $n_1$ ($z_{m_2, n_1} = 1, z_{m_2, n_2} = 0$). Since the calls are made close together in time, and $n_2$ is much closer to $m_1$ than $n_1$, detector $m_1$ receives the second call before the first. Without knowledge of which detection is of which call, we do not know the number of calls that generated the detected sounds. And without knowledge of the source locations of the calls, we don't even know the order in which the detected sounds are generated. This results in a combinatorial problem (which detected sounds came from which calls) with an unknown dimension (how many calls there are). 

The uncertainty in matching detections to make call capture histories can be seen as a latent identity (ID) SCR problem. Latent ID ASCR models in the literature have the individual identity as the latent variable, treating the call-level capture histories as known and only the identity of the individual associated with each capture history as unknown \citep[see][and references therein]{10.1093/biomtc/ujad019}. 
Here we treat the call identities as unknown and treat call capture histories as the latent variables. Whereas individuals may be detected multiple times by each detector, calls can be detected at most once by each detector and all detections by any one detector are necessarily of different calls. 
This is a similar situation to the two-camera aerial survey considered by \citet{https://doi.org/10.1111/biom.13403} and \citet*{https://doi.org/10.1111/biom.12983}, in which each camera can detect individuals at most once and all detections by any one camera are known to be of different individuals.

For the two-camera survey, \citet{https://doi.org/10.1111/biom.13403} proposed a maximum likelihood estimation (MLE)-based method that enumerates all possible matchings between detections, while \citet{https://doi.org/10.1111/biom.12983} fit an approximation to the Palm likelihood \citep*{https://doi.org/10.1002/bimj.200610339}. Unlike \citet{https://doi.org/10.1111/biom.13403}, our method addresses the uncertainty in capture history by sampling from the set of possible capture histories instead of enumerating all possible capture histories. We take advantage of the conditional independence between detections among different detectors (given call location and time), to develop a computationally efficient method that can deal with a large number of detectors with low computational cost.
Our method has some similarities with the spatial partial identity models (SPIMs) \citep{10.1214/17-AOAS1091} since both methods sample matchings. SPIM is a fully Bayesian MCMC inference method, whereas ours is a method of obtaining maximum likelihood estimates by sampling from, rather than marginalizing over, the latent identity variable.

Our inference method integrates a complete-data likelihood, a semi-complete data likelihood, and a conditional likelihood within a Monte Carlo expectation maximization (MCEM) framework \citep{563c04ff-76f3-3274-acb0-1393d981bfa4}. As a generalization of the EM algorithm, MCEM samples unobserved random variables conditioning on the observed data in the E-step and optimizing parameters in the M-step. It is a consistent estimator that will converge to the true MLE estimator when the iteration number goes to infinity \citep{10.1214/aos/1059655912}.
It is useful in the inference of complex statistical models where the E-step involves high-dimensional integration or where closed-form solutions are unavailable. In this application, we sample the unknown capture history in the E-step and estimate the unknown total number of calls $N$ in the M-step. To further decrease the computational cost, we partition detections into groups that could be detections of the same call, and between which no detections could be of the same call, and assemble the likelihood using these independent groups. 

We apply our method to the open-access \textit{A. lightfooti} dataset labeled by \citet{stevenson2015} using manual call matching. The dataset is challenging because these frogs emit a large number of calls within a brief temporal interval, which makes it difficult for human experts to match detections on different detectors as being of the same call. The dataset was relabeled by \citet{https://doi.org/10.1111/1365-2664.12810} using a different (apparently better) way of matching detections into call capture histories. ASCR inference with this revised dataset produces a density estimate that is substantially different (i.e. 40\%) from the estimate using the original call capture histories. For our analysis, we remove the matching information from the dataset. Using the frog survey as a basis for designing the simulation study, we test the point estimate and confidence interval estimate of our method.

We first express the automatic matching and density estimation problem with three likelihood models used for ASCR inferences in Section \ref{sec:ascr_likelihood}.
We then describe how to unite three likelihoods under the MCEM framework to address the problem in Section \ref{sec:mcem_detail}. We propose a graph-based algorithm to partition the detections into groups that are possibly from the same call to further decrease the computational cost in Section \ref{sec:like_parti}. At last, we introduce the way to generate interval estimates under our framework in Section \ref{sec:interval_esti}.

\section{ASCR Likelihoods}\label{sec:ascr_likelihood}

\subsection{Notation and Terminology}
We consider a survey with a duration $T$ in a survey region $A \subset {\rm I\!R}^2$  using $m \in \{1, ..., M\}$ detectors placed at known locations in $A$. The total number of calls within the survey area during the survey is $N$, which is unknown. We denote the latent capture history matrix as $\bm{Z}_{M \times N}$ with entry as $z_{m,n}$ indicating whether the call $n$ is detected by detector $m$. We can write $Z$ as a combination of row vectors: $\bm{Z} = (\bm{z}_{1,:}, ..., \bm{z}_{M,:})$, where $\bm{z}_{m,:} = (z_{m, 1}, ..., z_{m, N})$ is a binary vector indicating whether the calls $n \in \{1, ..., N\}$ are detected by detector $m$. $\bm{Z}$ matrix can also be written as a combination of column vectors $\bm{Z} = (\bm{z}_{:,1}, ..., \bm{z}_{:, N})$, where $\bm{z}_{:, n}$ indicates the binary state for the $n$th call across $M$ detectors. (This is what would usually be called the capture history for the $n$th call.)  Unlike the capture history commonly used in most SCR studies \citep{stevenson2015, borchers2008}, the capture history $\bm{Z}$ here is unknown, and we refer to it as a ``latent capture history''. One main difference between latent and observed capture histories is that latent capture histories allow empty entries, i.e. $\bm{z}_{:, n} = \bm{0}$, corresponding to a call that is missed entirely. The calls come from latent locations given by Cartesian coordinates $\bm{X} = (\bm{x}_1, ..., \bm{x}_{N})$, and the times of detections $\bm{T}$ come from latent call emisson times $\bm{e} = (e_1, ..., e_N)$.
  
Instead of the observed capture history, we observe $\bm{J} = (J_1, ..., J_M)$, the number of detections at each detector $m \in \{1, ..., M\}$. For detector $m$, the detection signal strengths are $\bm{y}_{m} = (y_{m, 1}, ..., y_{m, J_{m}})$, and the detection recording times are $\bm{t}_{m} = (t_{m, 1}, ..., t_{m, J_{m}})$. The signal strengths and detection times are organized in a time ascending order.
We combine the signal strength and recording time for all detectors to get $\bm{Y} = (\bm{y}_{1}, ..., \bm{y}_{M})$, $\bm{T} = (\bm{t}_{1},..., \bm{t}_{M})$. The signal strength $\bm{Y}$ and time of arrival $\bm{T}$ are unstructured data instead of structured matrices since the number of detections by each detector varies between detectors and we lack information about which detections on the different detectors are of the same call. 
Since the order in which calls are emitted is not necessarily the same as the order in which any one detector receives them, we denote the detection order for detector $m$ as $\bm{k}_m = (k_{m, 1}, ...,k_{m, j}, ..., k_{m, J_{m}})$, which is an element of the full permutation of a positive integer sequence $(1, ..., j, ..., J_{m})$. 
For example, in Figure \ref{fig:align_demo}, the detector $m_1$ has detection order as $\bm{k}_{1} = (2, 1)$ relative to $n_1, n_2$, and the detector $m_2$ has detection order as $\bm{k}_{2} = (1)$.
We can combine the detection orders for all detectors in the object $\bm{K} = (\bm{k}_{1}, ..., \bm{k}_{M})$.

However, conditioning on the latent capture history $\bm{Z}$ and detection order $\bm{K}$, the unstructured recorded signal strength $\bm{Y}$ and recorded time $\bm{T}$ can be padded and reordered into matrices $\tilde{\bm{Y}}_{M \times N}$ and $\tilde{\bm{T}}_{M \times N}$ with entries as $\tilde{y}_{m, n}$ and $\tilde{t}_{m, n}$. In detail, we have $\tilde{y}_{m,n} = \emptyset$ and $\tilde{t}_{m,n} = \emptyset$ when $z_{m, n} = 0$ since signal strength and detection time is only recorded when detection happens. And we have $\tilde{y}_{m, n_{{m,j}}} = y_{m, k_{m, j}}$ and $\tilde{t}_{m, n_{{m,j}}} = t_{m, k_{m, j}}$, where $k_{m, j} \in \bm{k}_{m}$ is the position of the detection in the ordered sequence of detections, and $n_{m,j}$ is the corresponding point in the subsequence of $n \in \{1, ..., N\}$ corresponding to successful detections at each detector (i.e. $z_{m, n_{m,j}} = 1$). For example in Figure \ref{fig:align_demo}, we have recorded signal strength as $\bm{Y} = \{(150, 133),(140)\}$, detection time as $\bm{T} = \{(2, 2.5),(1.7)\}$. Once we know the latent capture history 
$\bm{Z} = \left(\begin{array}{ccccc} 1, &1\\ 1, &0 \end{array}\right)$, and the detection order as $\bm{K} = \{(2,1),(1)\}$, we can have signal strength matrix $\tilde{\bm{Y}} = \left(\begin{array}{ccccc} 133, &150\\ 140, &\emptyset \end{array}\right)$ and detection time matrix $\tilde{\bm{T}} = \left(\begin{array}{ccccc} 2.5, &2\\ 1.7, &\emptyset \end{array}\right)$.

For simplicity, we do not usually show parameters as explicit arguments of the functions we develop and they will be introduced by the first time they are used.

\subsection{Complete-Data Likelihood}
We first consider the complete-data model in which the likelihood dimension is conditioned on the unknown number of calls, $N$, and assume all latent variables are observed. Then the likelihood has the form
\begin{longequation}
\begin{array}{ll}%

    L_{f} \propto & f(\bm{X} | N) f(\bm{e} | N) f(\bm{Z}, \bm{J}| \bm{X}) f(\bm{K} | \bm{X}, \bm{e}, \bm{Z}) \\ &  \times \ f(\bm{Y} | \bm{Z}, \bm{X}, \bm{K}) f(\bm{T} | \bm{Z}, \bm{X}, \bm{e}, \bm{K}) ,
     \label{equ:fcl}
         
\end{array}
\end{longequation}
where $\bm{e}$ is a vector of call emission times.

The equation is a generalized version of the Binomial model complete-data likelihood of \citet[Equation 21]{10.2307/24780847} for an acoustic survey with observed signal strength $\bm{Y}$, detection recording time $\bm{T}$, and call emission time $\bm{e}$. The main difference between our likelihood and theirs is that we include $\bm{K}$ to indicate the detection order on each detector. This is necessary for our application because, unlike \citet{10.2307/24780847}, we do not have data about the matching between detections (which calls are detected by which detectors). 
The detection order $\bm{K}$ is determined by the call locations $\bm{X}$, call emission times $\bm{e}$, and which calls are detected by which detectors $\bm{Z}$, so that the $f(\bm{K} | \bm{X}, \bm{e} , \bm{Z}) = 1$ for the $\bm{K}$ generated by $\bm{X}$, $\bm{e}$, and $\bm{Z}$, and $f(\bm{K} | \bm{X}, \bm{e} , \bm{Z}) = 0$ for all other $\bm{K}$. We will omit this probability density function (pdf) in the following content for simplicity.

We now consider each of the likelihood components. The pdf $f(\bm{X} | N)$ is the joint call location distribution conditioning on the number of calls produced within the survey ($N$): 
\begin{equation} \label{equ:f(x)}
    f(\bm{X} | N) = N! \prod_{n= 1}^{N}f(\bm{x}) = N! \prod_{n=1}^{N} \frac{\lambda(\bm{x})}{\Lambda} ,
\end{equation}
where $\lambda(\bm{x})$ is the animal intensity at the location $\bm{x}$ and $\Lambda = \int_{A}\lambda(\bm{x})d\bm{x}$ and there are $N!$ ways to get the same location set $\bm{X}$. 

$f(\bm{e} | N)$ is the joint call emission time distribution conditioned on $N$, which is treated as the product of the independent uniform distributions within a given time interval:
\begin{equation}
    f(\bm{e} | N) = \prod_{n= 1}^{N}f(e) = \prod_{n= 1}^{N} U(e_l, e_r),
\end{equation}
where $e_l, e_r$ are the survey start and end times. 

 $f(\bm{Z}, \bm{J}| \bm{X})$ is the latent capture history and detection count distribution conditioning on the latent call location. We assume that each call is detected independently by each detector, conditional on call location $\bm{X}$: 
\begin{equation}
    f(\bm{Z}, \bm{J} | \bm{X}) = \prod_{m =1}^{M}\prod_{n = 1}^{N} f(z_{m,n} | \bm{x}_n),
\end{equation}
where $J_m$, the number of calls detected by detector $m$, is a known function of $\bm{Z}$:  $J_m = \sum_{n = 1}^{N} z_{m,n}$, and $f(z_{m,n} | \bm{x}_n)$ is the pdf of the binary capture indicator for call $n$ at detector $m$ (see Web Appendix A for details).

 $f(\bm{Y} | \bm{Z}, \bm{X}, \bm{K})$ is the recorded signal strength pdf conditioning on the call location, capture history, and the detection order. After padding the signal strength into matrix $\tilde{\bm{Y}}$, we assume that the signal strengths are independent between each detector and each call, conditional on detection state $\bm{Z}$, detection order $\bm{K}$ and the call location $\bm{X}$:
\begin{equation}
    f(\bm{Y} | \bm{Z}, \bm{X}, \bm{K}) = f(\tilde{\bm{Y}}| \bm{Z}, \bm{X}) = \prod_{m=1}^{M} \prod_{n =1}^{N} f(\tilde{y}_{m,n} | z_{m,n}, \bm{x}_{n}),
\end{equation}
where $f(\tilde{y}_{m,n} | z_{m,n}, \bm{x}_{n})$ is the pdf of detected signal strength for call $n$ at detector $m$ (see Web Appendix A for details).
The detection order variable $\bm{K}$ is omitted from the RHS since the order information is already embedded in the padded signal strength matrix $\tilde{\bm{Y}}$.

 $f(\bm{T} | \bm{Z}, \bm{X}, \bm{e}, \bm{K})$ is the detection recording time distribution, conditional on the latent capture history, the latent call location, the latent call emission time, and the detection order across all detectors. Similar to the above, we can use the padded recording time matrix to substitute the original recording time: 
\begin{equation}
    f(\bm{T}| \bm{Z}, \bm{X}, \bm{e}, \bm{K}) = f(\tilde{\bm{T}}| \bm{Z}, \bm{X}, \bm{e}) =\prod_{m=1}^{M} \prod_{n=1}^{N} f(\tilde{t}_{m,n} | z_{m,n}, \bm{x}_n, e_{n}),
\end{equation}
where $f(\tilde{t}_{m,n} | z_{m,n}, \bm{x}_n, e_{n})$ is the pdf of recorded detection time for call $n$ at detector $m$ (see Web Appendix A for details).

\subsection{Semi-Complete-Data Likelihood} \label{sec:semi_complete}
We can split the data into two components according to the latent capture history $\bm{Z}$. That is, we split the data into observed and unobserved calls based on whether the calls have been detected at least once (i.e. $\bm{z}_{:,n} \neq \bm{0}$), where $\bm{X} = (\bm{X}^{o}, \bm{X}^{u})$, $\bm{Z} = (\bm{Z}^{o}, \bm{Z}^{u})$, $\bm{e} = (\bm{e}^{o}, \bm{e}^{u})$, $\tilde{\bm{Y}} = (\tilde{\bm{Y}}^{o}, \tilde{\bm{Y}}^{u})$, $\tilde{\bm{T}} = (\tilde{\bm{T}}^{o}, \tilde{\bm{T}}^{u})$ (Here superscript $^o$ indicates random variables for observed calls, while superscript $^u$ indicates random variables for unobserved calls). The number of calls, $N = N^{o} + N^{u}$, and the number of observed calls, $N^{o}$, are defined by
\begin{equation} \label{equ:cal_no}
    N^{o} = \sum_{n = 1}^{N}  \bmath{1} [\bm{z}_{:,n} \neq \bm{0}],
\end{equation}
where $\bmath{1}[\;]$ is the indicator function that takes value 1 when its argument is true, and 0 otherwise. 
We can write the semi-complete-data likelihood as follows:
\begin{equation} 
\begin{array}{ll}%
    L_{s}  \propto  & f(\bm{X}|N^{o}, N) f(\bm{e}^{o} | N^{o}) f(\bm{Z}^{o}, \bm{J}| \bm{X}^{o}) \\ & \times \
    f(\tilde{\bm{Y}}^{o} | \bm{Z}^{o}, \bm{X}^{o}) f(\tilde{\bm{T}}^{o} | \bm{Z}^{o}, \bm{X}^{o}, \bm{e}^{o}), 
\end{array}
    \label{equ:smi}
\end{equation}
where, conditional on the number of observed and unobserved calls, the pdf of call locations is 
\begin{equation}   \label{equ:f(X|No,N)}
    f(\bm{X}|N^{o}, N) =  \frac{N!}{N^{u}!}\prod_{n = 1}^{N^{o}} f(\bm{x}_{n}) p.(\bm{x}_{n}) \times \{ 1 - \int_{A}p.(\bm{x})f(\bm{x})d\bm{x} \} ^{N^{u}},
\end{equation}
where $p.(\bm{x}) = p(\bm{z} \neq \bm{0} | \bm{x})$ is the probability that a call has been detected at least once, and the $f(\bm{x}) = \frac{\lambda(\bm{x})}{\Lambda}$ is given in Equation (\ref{equ:f(x)}). (Since unobserved call locations are indistinguishable, the permutation number $N!$ is divided by $N^{u}!$.)

In addition, the pdf for the latent capture history and number of captures on each detector conditioning on the latent location of detected call $\bm{X}^{o}$ is 
\begin{equation}
    f(\bm{Z}^{o}, \bm{J}| \bm{X}^{o}) =  \prod_{n=1}^{N^{o}}\frac{\prod_{m=1}^{M}f(z_{m,n}| \bm{x}_{n})}{p.(\bm{x}_{n})}.
\end{equation}
Since now the latent capture history for each call has a non-zero entry (i.e., $\bm{z}_{:,n} \neq \bm{0}$), the pdf is conditioned on the call being detected at least once.  

The pdf of padded signal strength $\tilde{\bm{Y}}^{o}$, detection time $\tilde{\bm{T}}^{o}$, and call emission time $\bm{e}^{o}$ remains the same as that in the complete-data likelihood, but only for the detected calls.
Again the semi-complete-data likelihood model is generalized from the likelihood in \citet[Equation 24]{10.2307/24780847} with added $\tilde{\bm{Y}}^{o}, \tilde{\bm{T}}^{o}, \bm{e}^{o}$ for the acoustic survey and $\bm{K}$ for unknown detection order ($\bm{K}$ is omitted in the pdf since the information is embedded in $\tilde{\bm{Y}}^{o}$ and $\tilde{\bm{T}}^{o}$). Unlike the complete-data likelihood model, it integrates over the locations of unobserved calls (in Equation \ref{equ:f(X|No,N)}) instead of separately modeling each of these call locations.

\subsection{Conditional Likelihood}
When we assume the call location to be uniformly distributed (i.e., $f(x) = U(A)$), we can obtain a likelihood function conditional on the observed number of calls, $N^{o}$:  
\begin{equation} 
\begin{array}{ll}%
    L_{c} \propto  & f(\bm{X}^{o} | N^{o}) f(\bm{e}^{o} | N^{o}) f(\bm{Z}^{o}, \bm{J}| \bm{X}^{o}) \\ & \times \
    f(\tilde{\bm{Y}}^{o} | \bm{Z}^{o}, \bm{X}^{o}) f(\tilde{\bm{T}}^{o} | \bm{Z}^{o}, \bm{X}^{o}, \bm{e}^{o}),
    \label{equ:con}
\end{array}
\end{equation}
where the pdf for source locations of calls detected at least once is 
\begin{equation}
    f(\bm{X}^{o} | N^{o}) =  \prod_{n = 1}^{N^{o}} \frac{p.(\bm{x}_{n})U(A)}{\int_{A}p.(\bm{x})U(A)d\bm{x}} = \prod_{n = 1}^{N^{o}} \frac{p.(\bm{x}_{n})}{\int_{A}p.(\bm{x})d\bm{x}},
\end{equation}
and the pdf of latent capture history and detection count conditioning on the latent call location $\bm{X}^{o}$ for observed calls $f(\bm{Z}^{o}, \bm{J} | \bm{X}^{o})$ is the same as that in semi-complete-data likelihood. In addition, the pdf of recorded signal strength $f(\tilde{\bm{Y}}^{o} | \bm{Z}^{o}, \bm{X}^{o})$ and detection time $f(\tilde{\bm{T}}^{o} | \bm{Z}^{o}, \bm{X}^{o}, \bm{e}^{o})$ both remain the same. 

The details of each likelihood component are given in Web Appendix A. This includes
\begin{enumerate}

\item \textbf{the pdf of latent capture history} $z_{m,n}$, given call location $\bm{x}_n$: $f(z_{m,n}|\bm{x}_n)$, which depends on parameters $\bm{\gamma}=(\beta_0, \beta_1, \sigma_s)$; 

\item \textbf{the pdf of received signal strength} $\tilde{y}_{m,n}$, given the capture history $z_{m,n}$ and call location $\bm{x}_n$: $f(\tilde{y}_{m,n}|z_{m,n},\bm{x}_n)$, which depends on parameters $\bm{\gamma}=(\beta_0, \beta_1, \sigma_s)$;

\item \textbf{the pdf of detection time} $\tilde{t}_{m,n}$, given the capture history $z_{m,n}$, call location $\bm{x}_n$ and the call emission time $e_{n}$:  $f(\tilde{t}_{m,n}|z_{m,n}, \bm{x}_n, e_{n})$, which depends on a parameter $\bm{\phi} = (\sigma_t)$;

\item \textbf{the overall detection probability} $p.(\bm{x})$, given call location $\bm{x}$, which depends on parameters $\bm{\gamma}=(\beta_0, \beta_1, \sigma_s)$; 

\item \textbf{the pdf of latent call location} $\bm{x}$: $f(\bm{x})$, which depends on parameters $\bm{\psi}=(\bm{\beta}_s)$.
\end{enumerate}

\section{MCEM for Detection Matching} \label{sec:mcem_detail}
There are advantages and limitations to using each of the above likelihoods for inference. In the case of the complete-data likelihood, the capture history $\bm{Z}$, and the detection order $\bm{K}$ are conditionally independent between different detectors, given the latent call location $\bm{X}$ and call emission time $\bm{e}$. For example, we have $\bm{z}_{m_a,:} \perp\!\!\!\!\perp \bm{z}_{m_b,:} | \bm{X}, \bm{e}$ for $\forall{m_a, m_b} \in \{1,..., M\}, \  m_a \neq m_b$. This property allows us to sample $\bm{Z}, \bm{K}$ using the complete-data likelihood. Resampling capture histories is easy using this likelihood conditional on $N$, while $N$ itself is difficult to sample or estimate when $\bm{Z}$ and $\bm{K}$ are not known, whether one uses Bayesian MCMC methods (Bayesian data augmentation or reversible jump MCMC) or maximum likelihood estimation methods.

On the other hand, the number of calls, $N$, can be easily estimated from a semi-complete-data likelihood by maximum likelihood or from a conditional likelihood using a Horvitz– \\ Thompson-like estimator. But these likelihoods require the data to be separated into two parts based on the observation state (i.e., whether a call is observed or not, see Section \ref{sec:semi_complete} for details). This makes sampling the observed capture histories $\bm{Z}^{o}$ intractable because this matrix is constrained in both the row and column directions (i.e., $\sum \bm{z}_{m,:} = J_{m}$, $\sum \bm{z}_{:,n} > 0$). 

Here we use an MCEM algorithm to overcome the limitations of each of the above likelihoods, by sampling the latent variables from the complete-data likelihood in the algorithm's expectation step (E-step) and optimizing the semi-complete-data likelihood or the conditional likelihood in the maximization step (M-step). 
We show below that this E-step is equivalent to calculating the expected value of the semi-complete-data likelihood.
We also discuss briefly when it is preferable to use the semi-complete-data likelihood or the conditional likelihood in the M-step.

\subsection{MCEM algorithm}

The MCEM algorithm is an iterative method to find MLEs in statistical models with missing or hidden data \citep{563c04ff-76f3-3274-acb0-1393d981bfa4}. Let us use $\bm{u}$ to represent all latent variables (i.e. $\bm{X}$, $\bm{Z}$, $\bm{e}$, $\bm{K}$) and $\bm{o}$ to represent all observed data (i.e. $\bm{J}$, $\bm{Y}$, $\bm{T}$).  Let $\bm{\theta}$ represent all model parameters (i.e. $\bm{\gamma}$, $\bm{\phi}$, $\bm{\psi}$). The EM algorithm aims to obtain the MLE of $\bm{\theta}$ using some log-likelihood $\log L(\bm{\theta}; \bm{o})$ in an iterative way.
The MCEM algorithm alternates between the Monte Carlo E-step and the M-step. 
The E-step samples $\bm{u}_{1}, ..., \bm{u}_{S}$ from the associated conditional posterior $f(\bm{u} | \bm{o}; \bm{\theta}^{r}) = L(\bm{\theta}^{r}; \bm{u}, \bm{o}) / L(\bm{\theta}^{r}; \bm{o})$ using a Monte Carlo method such as Markov Chain Monte Carlo (MCMC) or importance sampling, where $\bm{\theta}^{r}$ is the estimate from $r \in \{1, ..., R\}$ MCEM iteration. These samples are then used to estimate the expected log-likelihood
\begin{equation}
    Q(\bm{\theta} | \bm{\theta}^{r})  = \frac{1}{\Delta} \sum_{\delta = 1}^{\Delta} log L(\bm{\theta} ; \bm{u}_{\delta}, \bm{o}) \xrightarrow{P} E_{\bm{u}| \bm{o}; \bm{\theta}^{r}}[log L(\bm{\theta}; \bm{u}, \bm{o}) ],
\end{equation}
where $\Delta$ is the number of MC samples. $Q(\bm{\theta} | \bm{\theta}^{r})$ is called the evidence lower bound (ELBO) of the original likelihood $L(\bm{\theta}; \bm{o}) = \int_{\bm{u}}L(\bm{\theta}; \bm{u}, \bm{o})d\bm{u}$ \citep{543975}. The M-step maximizes this expected log-likelihood with respect to the parameters, to obtain updated parameter estimates:
\begin{equation}
    \bm{\theta}^{r+1} = \mathop{\arg\max}\limits_{\bm{\theta}} \ Q(\bm{\theta} | \bm{\theta}^{r}).
\end{equation}
The MCEM algorithm repeats the Monte Carlo E-step and M-step iteratively until convergence.

\subsection{Application of MCEM to ASCR Model}
We utilise the ELBO function according to the semi-complete-data likelihood:
\begin{equation}
    Q_s(\theta | \theta^{r}) = E_{\bm{u}^{s} | \bm{o}; \bm{\theta}^{r}}[log L_{s}(\bm{\theta}; \bm{u}^{s}, \bm{o})] .
\end{equation}
However, one cannot directly compute the $Q_s(\theta | \theta^{r})$ in MCEM E-step because it is intractable to sample $\bm{u}^{s}$ from the semi-complete-data likelihood model, with the constraints on both rows and columns of the latent capture history matrix. On the other hand, one could easily sample from the complete-data likelihood instead, and we define the new ELBO function as
\begin{equation}
    Q_{fs}(\theta | \theta^{r}) = E_{\bm{u^{f}} | \bm{o}; \bm{\theta}^{r}}[log L_{s}(\bm{\theta}; \bm{u}^{s}, \bm{o})] ,
\end{equation}
where $\bm{u}^{f}$ is the latent variable within the complete-data likelihood model. 
And immediately we have (see Web Appendix B for details)
\begin{equation}
    Q_{fs}(\theta | \theta^{r}) =  Q_s(\theta | \theta^{r}) ,
\end{equation}
which is the key property for using MCEM in ASCR inference. 
This allows us to sample from the complete-data likelihood in the E-step while optimizing the semi-complete-data likelihood model in the M-step.

We can also define the ELBO function as
\begin{equation}
    Q_{fc}(\bm{\theta} | \bm{\theta}^{r}) = E_{\bm{u}^{f}| \bm{o}; \bm{\theta}^{r}}[log L_{c}(\bm{\theta}; \bm{u}^{s}, \bm{o})] , 
\end{equation}
where the semi-complete-data likelihood $L_s$ is substituted with the conditional likelihood $L_c$.
\citet{10.1093/biomet/71.1.27} showed that MLEs of $\bm{\theta}$ from the conditional and semi-complete-data likelihood functions are equivalent. We then have
\begin{equation}
    \bm{\theta}^{r+1} = \mathop{\arg\max}\limits_{\bm{\theta}} \ Q_{fs}(\bm{\theta} | \bm{\theta}^{r}) = \mathop{\arg\max}\limits_{\bm{\theta}} \ Q_{fc}(\bm{\theta} | \bm{\theta}^{r}) . 
\end{equation}
This allows us to move between complete-data likelihood and conditional likelihoods during the MCEM procedure.

\subsection{Monte Carlo E-Step}
During the E-step, we need to sample latent variables $\bm{X}, \bm{Z}, \bm{e}, \bm{K}$ in an iterative way. Based on the complete-data likelihood in Equation (\ref{equ:fcl}), we have the sampling posterior for $\bm{Z}$ and $\bm{K}$ as $f(\bm{Z}, \bm{K} | \bm{X}, \bm{J}, \bm{Y}, \bm{T}, \bm{e})$. We sample $\bm{Z}$ and $\bm{K}$ together since $\bm{K}$ is determined by $\bm{Z}$.
For the latent capture history and detection order, we have the full conditional 
\begin{equation}
    f(\bm{Z}, \bm{K} | \bm{X}, \bm{J}, \bm{Y}, \bm{T}, \bm{e}) \propto  f(\bm{Z}, \bm{J}| \bm{X}) f(\bm{Y} | \bm{Z}, \bm{X}, \bm{K}) f(\bm{T} | \bm{Z}, \bm{X}, \bm{e}, \bm{K}).
\end{equation}
Since the capture histories and the detection order across different detectors are independent conditional on the latent call location $\bm{X}$ and call emission time $\bm{e}$, we obtain the sampling posterior for a single detector $m$ as
\begin{longequation}
\begin{array}{ll}%
    f(\bm{z}_{m, :}, \bm{k}_m | \bm{X}, J_{m}, \bm{y}_{m}, \bm{t}_{m}, \bm{e}) \propto  f(\bm{z}_{m, :}, J_{m}| \bm{X}) f(\bm{y}_{m} | \bm{z}_{m, :}, \bm{X}, \bm{k}_m)  f(\bm{t}_{m} | \bm{z}_{m, :}, \bm{X}, \bm{e}, \bm{k}_m),
\end{array}
\end{longequation}
where the summation of latent call capture history $\bm{z}_{m, :}$ at detector $m$ is constrained by the detection number $J_{m}$. This posterior is still intractable since obtaining the normalizing constant is computationally intractable. We therefore use a Metropolis-Hastings (MH) algorithm to sample these latent variables.

The sampling posterior for the latent location is 
\begin{equation}
   f(\bm{X} | \bm{Z}, \bm{Y}, \bm{T}, \bm{e}, \bm{K}) \propto f(\bm{X}) f(\bm{Z} | \bm{X}) f(\tilde{\bm{Y}} | \bm{Z}, \bm{X}) f(\tilde{\bm{T}} | \bm{Z}, \bm{X}, \bm{e}).
\end{equation}
We omit $\bm{J}$ from the sampling posterior since the latent location is conditionally independent of the number of detections when a latent capture history exists.
Since the prior for the  latent location of each call is assumed to be independent, we have the sampling posterior for each individual call location as 
\begin{equation}
f(\bm{x}_{n} | \bm{z}_{:,n}, \tilde{\bm{y}}_{:, n}, \tilde{\bm{t}}_{:,n}, e_{n}) \propto f(\bm{x}_n) f(\bm{z}_{:,n} | \bm{x}_{n}) f(\tilde{\bm{y}}_{:, n} |  \bm{z}_{:, n}, \bm{x}_n) f(\tilde{\bm{t}}_{:,n} |  \bm{z}_{:, n}, \bm{x}_n, e_{n}),
\end{equation}
where the detection order $\bm{K}$'s information is embedded in $\tilde{\bm{Y}}$ and $\tilde{\bm{T}}$.
The posterior for each call location is still intractable so we use an MH algorithm to sample the location $\bm{x}_{n}$. 

Similarly, the sampling posterior for the latent call emission time is
\begin{equation}
        f(\bm{e} |\bm{X}, \bm{Z}, \bm{T}, \bm{K}) \propto  f(\bm{e})  f(\tilde{\bm{T}} | \bm{Z}, \bm{X}, \bm{e}).
\end{equation}
Since the prior for the call emission time is assumed to be independently uniformly distributed, and the call detection order $\bm{K}$'s information is included in $\tilde{\bm{T}}$ matrix, we have the posterior for individual call emission time as
\begin{equation}
    f(e_{n}| \bm{x}_n, \bm{z}_{:,n}, \tilde{\bm{t}}_{:,n}) \propto f(e_{n}) f(\tilde{\bm{t}}_{:,n} |  \bm{z}_{:, n}, \bm{x}_n, e_{n}).
\end{equation}
Since the prior $f(e_{n})$ is uniformly distributed, and the $f(\tilde{\bm{t}}_{:,n} |  \bm{z}_{:, n}, \bm{x}_n, e_{n})$ is a product of independent Normal distributions at each detector, the posterior is a truncated Normal distribution, and we can sample $e_{n}$ directly from this. The details of the sampling procedures in the E-step are given in Web Appendix C.

\subsection{M-Step}
During the M-step, if we assume an independent non-homogeneous spatial Poisson point process for the locations of detected calls, we can use the semi-complete-data likelihood model in optimization, where the estimated parameter vector is $\bm{\theta}^{r+1} = ({N}^{r+1}, \bm{\gamma}^{r+1}, \bm{\phi}^{r+1}, \bm{\psi}^{r+1})$.
However, each individual might emit calls more than once during the survey and this introduces dependence on call locations. When detected call locations are not independent, we could use the conditional likelihood with a uniform assumption on call location instead (i.e. $f(\bm{x}) = U(\bm{A})$) \citep{stevenson2015}. This allows us to estimate an average calling density $D_c$ across the survey area and time but not the spatial intensity function $\lambda(x)$. In this case, assuming the non-independent variables to be independent will strongly affect interval estimation, although point estimation is not necessarily biased by the assumption \citep{stevenson2015}. 

Maximizing the ELBO of conditional likelihood (\ref{equ:con}) with respect to the parameters using sampled $\bm{X}, \bm{Z}, \bm{K}, \bm{e}$ from E-step, we estimate $\bm{\theta}^{r+1} = (\bm{\gamma}^{r+1}, \bm{\phi}^{r+1})$.
We can then estimate the mean detection probability $p$ by evaluating it at the estimates of these parameters with regard to the maximized ELBO function:
\begin{equation}
    p^{r+1} = \frac{\int_{A} p.^{r+1}(\bm{x}) d \bm{x}}{A},
\end{equation}
where $A$ is the area of the survey region, and $p.(\bm{x})$ is the probability that a call at location $\bm{x}$ is detected at all. 
Density is then estimated by
\begin{equation}
    D^{r+1}_{c} = \frac{N^{o}}{p^{r+1}  A  T} ,
\end{equation}
where $N^{o}$ is the number of detected calls calculated from sampled $\bm{Z}$ using Equation (\ref{equ:cal_no}) and $T$ is the survey duration.
Using this, we can estimate the total number of calls as follows:
\begin{equation}
    N^{r+1} = D^{r+1}_c \times A \times T .
\end{equation}

It is worth noting that we can use a slightly different conditional likelihood in M-step, with $N^{o}$ defined by the number of calls being detected by at least two detectors. The relevant likelihood in this case retains the same structure as shown above but uses an overall detection probability equal to $p.(\bm{x}) = p(\sum \bm{z} > 1 | \bm{x})$ \citep[see][and Web Appendix A for details]{https://doi.org/10.48550/arxiv.2207.09343, https://doi.org/10.1111/2041-210X.14326}. Discarding sampled calls that are detected by only one detector, which tends to have source locations far away from the array, removes calls with the least distinguishable latent call locations and makes the M-step more stable.

\section{Graph-Based Likelihood Partition} \label{sec:like_parti}
Some detections are too far away in time for them to possibly be detections of the same call. Because sampling latent capture histories of all detections together can be very computationally expensive, we use a graph-based pre-processing method to separate detections that are too far apart in time to be detections of the same call.
We partition all observed data $\bm{o} = (\bm{J},\bm{Y}, \bm{T})$ by their recorded time $\bm{T}$ using a threshold-based algorithm. More specifically, we represent detections by an undirected graph $G = (V, E)$, where vertices set $V$ include all of the individual detections (detections of a single sound at a single detector) $o \in V$, and the edge set $E$ includes pairs of distinct vertices $E \subseteq \{ (o_{a}, o_{b}) | (o_{a}, o_{b}) \in V^{2} \ and \ o_{a} \neq o_{b} \}$. We define edge to be when (1) two detections $o_{a}$ and $o_{b}$ are not from the same detector $m_{a} \neq m_{b}$; (2) the absolute time difference between two detections at two different detectors $m_a, m_b$ is smaller than the theoretical maximum time difference (i.e. the time it takes sound moving at $v$=330 m/s to travel the distance $d_{a,b}$  between the two detectors) plus an amount $3\sigma$ to account for variation in the speed sound travels and recording time error: 
\begin{equation}
    |t_{a} - t_{b}| < d_{a,b} / v + 3\sigma.
\end{equation}
Here $\sigma$ is a hyper-parameter that should be decided based on the prior knowledge of the recording device and survey setup. If $\sigma$ is too small, some recaptures of the same call may be treated as different calls; if it is too large, this will slow computation.

After deciding the edge and vertices of the graph, we can partition the graph $G$ to connected sub-graphs $G_{1}, ..., G_{I}$, where all vertices pairs within the same sub-graph are connected (i.e. starting from one vertex, edges exist that eventually lead to the another), while any vertex pairs chosen from different sub-graphs are not connected. Then we can obtain the likelihood as
\begin{equation}
    L(\bm{\theta}; \bm{o}) = \prod_{i=1}^{I}L(\bm{\theta}; \bm{o}^{i}),
\end{equation}
where $\bm{o}^{i}$ is the observed data in sub-graph $G_{i}$.

\section{Interval Estimation} \label{sec:interval_esti}
MCEM is typically used only for point estimation. However, \citet{10.1214/08-BA326} points out that interval estimation can be achieved with an expected information matrix proposed by \citet{2c2e2dca-a57f-3d94-a89e-09a35ef7225a}:
\begin{equation}
    I_{\bm{o}} = I(\hat{\bm{\theta}}) = I_{\bm{u}, \bm{o}} - I_{\bm{u}| \bm{o}},
\end{equation}
where $I_{\bm{u} | \bm{o}} = E_{\bm{u} | \bm{o}; \hat{\bm{\theta}}}[J(\bm{u}, \bm{o}, \hat{\bm{\theta}})J^{T}(\bm{u}, \bm{o}; \hat{\bm{\theta}})]$ is the expectation of the squared likelihood Jacobian matrix, and $I_{\bm{u}, \bm{o}} = E_{\bm{u} | \bm{o}; \hat{\bm{\theta}}}[H(\bm{u}, \bm{o}, \hat{\bm{\theta}})]$ is the expectation of the likelihood Hessian matrix.

When the detected call locations are not independent, the Fisher information matrix will not be useful for interval estimation \citep{stevenson2015}. Following \citet{doi:10.1080/02664763.2013.825704}, we apply a bootstrap method to estimate the variance of the parameters in this case, except we use a parametric bootstrap instead of a non-parametric bootstrap. During the bootstrap, instead of running $R$ iterations of the MCEM schedule for each bootstrap sample, we only do a few iterations (1 iteration in the extreme case) of the E-step and M-step, where we initialize $\bm{\theta}$ as $\hat{\bm{\theta}}_{init} = \mathop{\arg\max}\limits_{\bm{\theta}} L(\bm{\theta}; \bm{u}^{*}, \bm{o}^{*})$.  $\bm{u}^{*}$ and $\bm{o}^{*}$ are both known in each bootstrap sample since we simulate them. Here $\hat{\bm{\theta}}_{init}$ is close to the MLE of $\hat{\bm{\theta}} = \mathop{\arg\max}\limits_{\bm{\theta}} L(\bm{\theta}; \bm{o}^{*})$ in the MCEM schedule so that the algorithm requires fewer steps to converge. This schedule decreases the computational time of the bootstrap, thus increasing the method's efficiency. The details of the bootstrap algorithm can be found in Web Appendix D.

\section{Frog Survey Analysis}
We applied our method to the moss frog (\textit{A. lightfooti}) survey data published by \citet{https://doi.org/10.15493/saeon.metacat.10000005}. The data were recorded by 6 detectors on 16 May 2012, and 25 seconds of the recordings were manually labeled and matched (i.e., decisions made manually about which detections are of the same call) by \citet{stevenson2015}. The detections were re-matched by \citet{https://doi.org/10.1111/1365-2664.12810} to produce what are considered to be more reliable call capture histories. For simplicity, we refer to these as ``manual matching ASCR (v1)'' and ``manual matching ASCR (v2)'' in the following text.
In this study, the call locations can not be regarded as independent since an individual may produce more than one call during the survey. We, therefore, use the conditional likelihood and not the semi-complete-data likelihood in the M-step. We use 1000 bootstrap samples to generate interval estimates. 
We compare our call density estimate with the result generated by ASCR with two versions of manual matching (see Table \ref{tab:compare_esti}).

\begin{table}
\caption{Point estimates, coefficient of variation (CV), and confidence interval (CI) of frog call density $D_c$ by manual matching ASCR (v1), manual matching ASCR (v2), and our automated matching ASCR methods. Estimated units are quantities per second per hectare}
\label{tab:compare_esti}
\begin{center}
\begin{tabular}{lccc}
\Hline
Methods                  &\textit{Estimates} & \textit{CV (\%)} &\textit{CI} \\ 
\hline
Manual matching ASCR (v1)     & 99.15 & 17.26   & (67.46, 136.24)      \\
Manual matching ASCR (v2)  & 60.99   & 18.60  & (40.54, 86.04)     \\ 
Automated matching ASCR    & 52.35  & 26.67   & (30.24, 88.00)   \\
\hline
\end{tabular}
\end{center}
\end{table}

After removing the matching information from the dataset, we managed to achieve the call density estimate with separate detections on each detector and their recorded time and signal strength. 
Our estimate is within 15\% of the estimate from the more reliable capture histories (ASCR (v2)), and much closer to this estimate than that from the initial capture history dataset (ASCR (v1)). The coefficient of variation (CV) from our method is larger than those from both of the manual matching methods. This is unsurprising since our estimator reflects the uncertainty about capture histories, while the ASCR (v1) and ASCR (v2) estimates assume no uncertainty in detection matching. The assumption of no uncertainty in capture histories when doing manual matching seems unrealistic to us, and we note that the two density estimates from the two sets of manual matchings differ by almost 40\%. The CV is calculated as $\frac{\hat{\sigma}(\hat{D}_{c}^{*})}{\hat{\mu}(\hat{D}_{c}^{*})}$, where $\hat{\sigma}(\hat{D}_{c}^{*})$ and $\hat{\mu}(\hat{D}_{c}^{*})$ are standard deviation and mean of the bootstrap samples. The confidence interval (CI) is calculated using the percentile method \citep{davison_hinkley_1997}. Our CI estimate covers the CI estimate of manual matching ASCR (v2), suggesting some consistency in point estimation between estimates from the more reliable manual matching method and our method.

\section{Simulation} \label{sec:simulation}
We investigate the properties of our estimator by parametric simulation, following \citet{stevenson2015}. We generated 1000 datasets independently, where the values used for the simulation parameters are set at the corresponding estimates obtained from the real data analysis, except we put the detectors in the array slightly further apart from each other than in the real survey. Setting the detectors further apart will increase the difficulty in detection matching since it takes a longer time for sound to travel from one detector to another thus allowing more candidate detections to match with each other (see Section \ref{sec:like_parti} for details).
On the other hand, we found that when the detectors are set too close to each other, latent call locations become less distinguishable. (This issue will also apply when doing manual matching and is a consideration that should be taken into account when designing a survey.)

Confidence interval coverage is tested using 200 randomly chosen subsets of the 1000 generated datasets to prevent extremely long simulation time. 

\begin{figure}[]   
         \centering
         \includegraphics[width=6.5in]{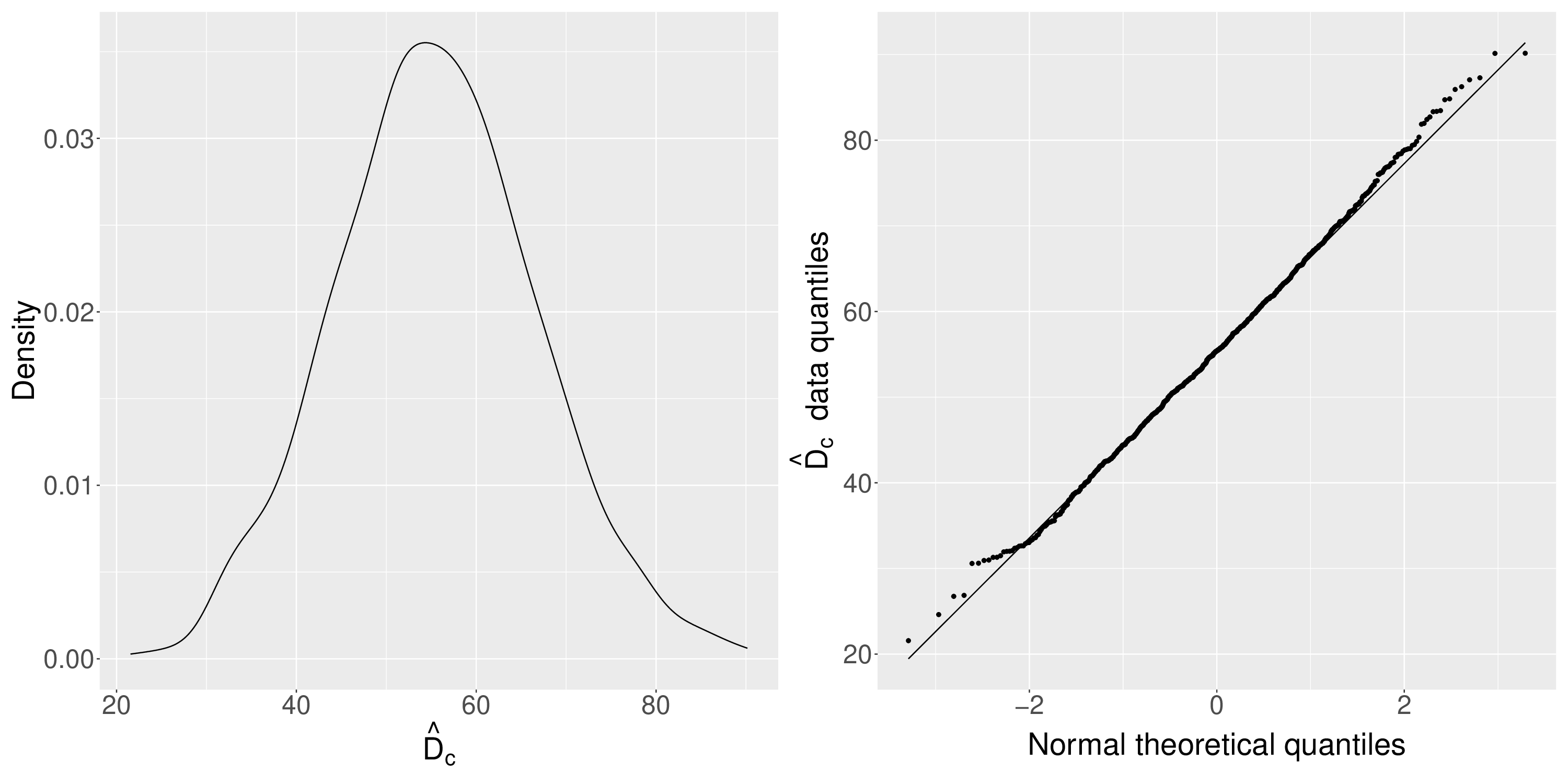}
          \caption{The density plot and Normal QQ plot for simulated $D_c$ estimation.}
         \label{fig:dist_dc}
\end{figure}

\begin{figure}[]   
         \centering
         \includegraphics[width=6.5in]{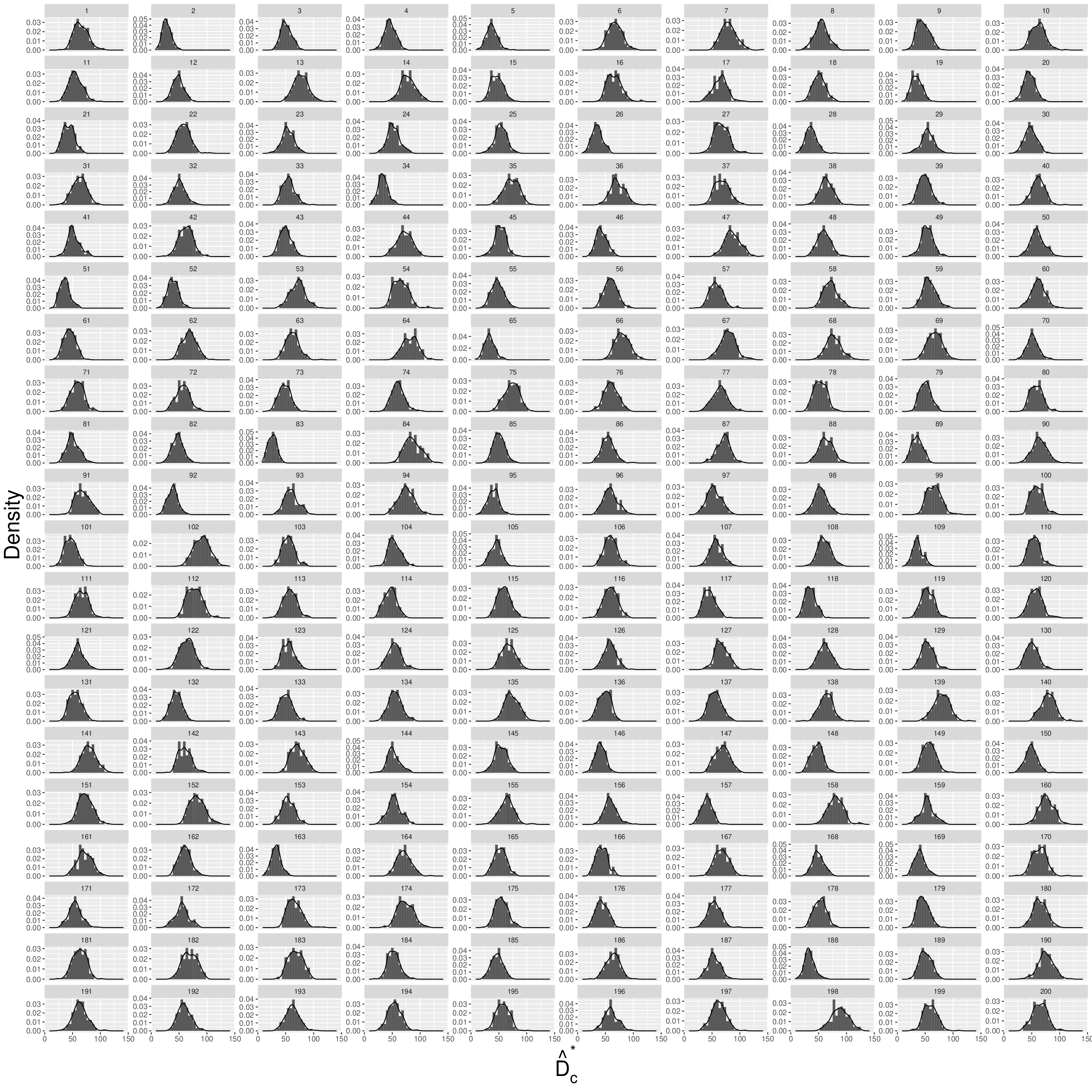}
         \caption{The density plot of estimated $D_c$ from 200 independent bootstrap simulations.
         }
         \label{fig:boot_dc}
\end{figure}

We found our method to have a low bias (6.07 \%, compared to a simulated CV of 20.13\%) and coverage close to the nominal 95\% probability (i.e. 93\%). The density plot and Normal QQ plot indicate approximate normality but the distribution is slightly right-skewed
(see Figure \ref{fig:dist_dc} for details).
We suggest calculating the confidence interval based on the \textit{percentile} method instead of assuming perfect normality. The density estimate from bootstrap sampling shows similar right skewness as the simulation distribution (see Figure \ref{fig:boot_dc} for details).

\section{Discussion}

\citet{https://doi.org/10.1111/biom.13403} developed a maximum likelihood method for dealing with recapture uncertainty when there are only two detectors, but it requires enumeration of all possible capture histories. With just two detectors, the number of possible capture histories is $N_{ch} = \sum_{i = 0}^{J_1}{J_1 \choose i}{J_2!}/{(J_2 - i)!}$ (where $J_1$ and $J_2>J_1$ are the numbers of detections by each detector.) As the detection number $J_2$ grows, enumerating all possible capture histories quickly becomes computationally infeasible. When there are $M>2$ detectors in the survey, the number of possible capture histories becomes intractably large, even if each detector only has a few detections.
\citet{https://doi.org/10.1111/biom.12983} developed a computationally efficient likelihood approximation method for the two detector cases, and this remains efficient even when $J_1$ and $J_2$ are large, but it does not generalize easily to the case of more detectors. By exploiting the conditional independence between detectors, 
our method can cope with more detectors (6 in the application above), and remain computationally feasible. (It generated density estimates within minutes in our application.) Full Bayesian methods such as \citet{10.1214/17-AOAS1091} developed in SPIMs can not be easily generalized to our case since call locations can have dependence, and treating non-independent call locations as independent greatly affects the interval estimates. We overcome this by using a parametric bootstrap that builds the dependence into the bootstrap to estimate confidence intervals. The bootstrap procedure can be time-consuming because the inference on many bootstrap samples costs a significant amount of time. This can be mitigated by parallel computation as bootstrap samples are conditional independent of each other.

We found our estimator to have a small positive bias in the scenario we simulated. We postulate that this may result from the way we partition detections into small groups within which detections possibly come from the same call. The number of detections $J$ on each detector can be seen as a Poisson-binomial random variable conditioned on $N$ with pdf $f(J|N)$, and partitioning into groups violates the dependency between detections. That is, we assume that $f(J|N) = \prod_{i=1}^{I}f(J^{i}|N^{i})$ during likelihood partition (see Section \ref{sec:like_parti}). Small positive bias was also found by \citet{https://doi.org/10.1111/biom.13403},  which uses a similar detection partitioning method. On the other hand, partitioning detections into groups can significantly decrease the computational cost.

During simulation, we found that placing detectors too close to each other negatively affects the convergence robustness of ASCR methods in general and affects our method in particular because the closer together the detectors, the less information there is on call location, and so the less information there is on which detections are of the same call.
We suggest using detectors with a variety of spacings and using greater separation for species whose calls can be heard farther away. Using sound features of individual detections is another way to make the detection more distinguishable, although this would require an extension of our method. We expect that this extension will not be overly difficult because our method uses ASCR likelihoods explicitly, so the development of an ASCR likelihood that includes sound feature information is really all that is required.

A core idea underpinning the paper is that with ASCR, one can use different kinds of likelihood in the same MCEM framework. We demonstrate that this is valid when one likelihood is the marginal likelihood with some latent variables integrated out (the semi-complete-data likelihood is the complete-data likelihood with some of the latent variables integrated out). The M-step is also changeable between models that share the same MLE estimate; e.g., the semi-complete-data likelihood and the conditional likelihood in our case. 
By sampling from one likelihood in the Monte Carlo E-step and optimizing a marginal or conditional likelihood in the M-step, we are able to deal with the difficulties posed by the fact that the complete-data likelihood has an unknown dimension. Similar to the idea of \citet*{10.1111/1467-9868.00179}, the auxiliary random variables are used for sampling simplicity, although the sampled variables are then used for ELBO optimization instead of MCMC sampling. This method may be applicable to other ecological surveys when the detection ID is unknown.









\backmatter


\section*{Acknowledgements}
YW is funded by the China Scholarship Council (CSC) (Grant Number 202008060348) for Ph.D. study at the University of St Andrews, UK. \\
For the purpose of open access, the author has applied a Creative Commons Attribution (CC BY) licence to any Author Accepted Manuscript version arising. \vspace*{-8pt}

\section*{Data Availability Statement}
The data used in this paper to illustrate our findings are available in the GitHub \textit{ascr} package, at https://github.com/b-steve/ascr.


%
 \bibliographystyle{biom} 
\bibliography{autodetscr}








\newpage
\appendix


\section*{Web Appendix A - Detail of the Likelihood Components}

\subsection*{The pdf of Latent Capture History} \label{Appen:pdf_bincapt}
For $n$th call detected at detector $m$, the capture history $z$ is a Bernoulli random variable with $g\{d_{m}(\bm{x}_n)\}$ as parameter, where $d_{m}(\bm{x}_n)$ is the distance between call location $\bm{x}_n$ and detector $m$. $g(d)$ is the detection probability as a function of the distance:
\begin{equation} \label{equ:det_auto}
    g(d) = \int_{-\infty}^{\infty} p(z = 1 | y) f(y | d)dy ,
\end{equation}
where $p(z = 1 | y)$ is the detection probability given signal strength $y$, which is usually modeled as a step function or a monotonically increasing function between 0 and 1 \citep{wang2023automated, stevenson2015}.
Following \citet{https://doi.org/10.1890/08-1735.1} and \citet{stevenson2015}, we assume $f(y|d)$ to be a Normal distribution with variance $\sigma^2_s$ and a mean that is a monotonically declining function of the distance $d$ between call and detector, with parameters  $\beta_0, \beta_1$ :
\begin{equation}
     f(y | d) = N(y| E[y|d; \beta_0, \beta_1], \sigma_{s}^2).
\end{equation}
We applied a linear attenuation function in this application:
\begin{equation} \label{equ:decay_fp}
E[y|d; \beta_0, \beta_1] = \beta_0 -  \beta_1 \times d  , 
\end{equation}
where $\beta_0$ is the source signal strength, $\beta_1$ is the linear attenuation of the signal strength, and $d$ is the distance between the call and the detector. Other forms of sound attenuation functions are also applicable to our method.

\subsection*{The pdf of Received Signal Strength} \label{Appen:pdf_ss}
Since signal strength $y$ is only recorded if a call is detected, we model the observed signal strength conditional on detection ($z=1)$ using the Bayes' rule:
\begin{equation} \label{equ:ss_auto}
    f(y | d, z =1) = \frac{p(z = 1 | y) f(y | d )}
    {\int_{-\infty}^{\infty} p(z = 1 | y) f(y | d)dy},
\end{equation}
\noindent
where the detection probability function $p(z = 1| y)$ and the pdf of signal strength $y$ conditioning on the distance $f(y | d)$ have been defined in the above.
And we define 
\begin{equation} 
    f(y | d, z =0) = 1 ,
\end{equation}
when the binary detection state $z$ is 0.

\subsection*{The pdf of Detection Time} \label{Appen:pdf_toa}

The pdf of detection time is assumed to be a Gaussian distribution conditioning on the distance $d$ between call location $\bm{x}$ and detector, and call emission time $e$, with variance $\sigma_t$ as the parameter:
\begin{equation} \label{equ:toa_pdf}
    f(t | d, e, z = 1) = N(t| e + d/v, \sigma_t),
\end{equation}
where $v$ is the sound speed.
Since the detection time is only recorded when a call is detected (i.e. $z = 1$), we define  
\begin{equation} 
    f(t | d, e, z =0) = 1 ,
\end{equation}
when a call is not detected by a detector.

\subsection*{The Overall Detection Probability} \label{Appen:pdf_loc}
The overall detection probability at a given location is assumed to be
\begin{equation}
    p.(\bm{x}) = 1 - \prod_{m=1}^{M} 1 - g\{d_m(\bm{x})\},
\end{equation}
where $g(d)$ is the probability of detection with given distance $d$. This is the probability that a call is detected at least once. However, one could also define this to be the probability that at least two detectors detect the call \citep{https://doi.org/10.48550/arxiv.2207.09343}:
\begin{equation}
    p.(\bm{x}) = 1 - \prod_{m=1}^{M} 1 - g\{d_m(\bm{x})\} - \sum_{m_i=1}^{M} g\{d_{m_i}(\bm{x})\} \prod_{m_j \neq m_i} 1 - g\{d_{m_j}(\bm{x})\}.
\end{equation}

\subsection*{The pdf of Latent Call Location}
When individual location can be seen as independent, one could assume a non-homogeneous spatial Poisson point process with intensity \citep{10.2307/24780847}
\begin{equation}
    \lambda(\bm{x}) = e^{\bm{\beta}_s \bm{x} } , 
\end{equation}
and we have 
\begin{equation}
    f(\bm{x}) = \frac{\lambda(\bm{x})}{\int_A \lambda(\bm{x})d\bm{x}} ,
\end{equation}
where $A$ is the survey region.
However, in this application, we assume 
\begin{equation}
    f(\bm{x}) = U(A) .
\end{equation}
This allows us to estimate the overall call density but not the Poisson intensity function.

\vspace{3em}

\section*{Web Appendix B - Proof of the Equivalence of the ELBO Function}

According to Section \ref{sec:semi_complete}, the semi-complete-data likelihood is the complete-data likelihood integrated over the locations of the unobserved calls (thus all the random variables related to unobserved calls, which we define as $\bm{u}^{h}$) \citep{10.2307/24780847}. 
We then have $L_{s}(\bm{\theta}; \bm{u}^{s}, \bm{o}) = \int_{\bm{u}^{h}}L_{f}(\bm{\theta}; \bm{u^{f}}, \bm{o})d\bm{u}^{h}$ and $\bm{u^{f}} = \bm{u}^{s} \cup \bm{u}^{h}$. Thus, we immediately have 
\begin{longequation}
\begin{array}{ll}%
     Q_{fs}(\bm{\theta} | \bm{\theta}^{r})  
    &= E_{\bm{u^{f}}| \bm{o}; \bm{\theta}^{r}}[log L_{s}(\bm{\theta}; \bm{u}^{s}, \bm{o})]  
    \\&= E_{\bm{u^{s}, \bm{u}^{h}}| \bm{o}; \bm{\theta}^{r}}[log L_{s}(\bm{\theta}; \bm{u}^{s}, \bm{o})]  
    \\&= E_{\bm{u^{s}}| \bm{o}; \bm{\theta}^{r}}[log L_{s}(\bm{\theta}; \bm{u}^{s}, \bm{o})]  
    \\ &= Q_s(\bm{\theta} | \bm{\theta}^{r})  .
\end{array}
\end{longequation}

\vspace{3em}

\section*{Web Appendix C - Detail of the MCMC Sampling in E-Step}

\subsection*{Sampling Detail of Latent Capture History $\bm{Z}$ and Detection Order $\bm{K}$}

For the latent capture history $\bm{Z}$ and detection order $\bm{K}$, we have the full conditional
\begin{equation}
    f(\bm{Z}, \bm{K} | \bm{X}, \bm{J}, \bm{Y}, \bm{T}, \bm{e}) \propto  f(\bm{Z}, \bm{J}| \bm{X}) f(\bm{Y} | \bm{Z}, \bm{X}, \bm{K}) f(\bm{T} | \bm{Z}, \bm{X}, \bm{e}, \bm{K}) .
\end{equation}
Since the capture histories and the detection order across different detectors are independent conditional on the latent call location $\bm{X}$ and call emission time $\bm{e}$, we can obtain the sampling posterior for a single detector $m$ as
\begin{equation}
\begin{array}{ll}%
    f(\bm{z}_{m, :}, \bm{k}_m | \bm{X}, J_{m}, \bm{y}_{m}, \bm{t}_{m}, \bm{e}) \propto  f(\bm{z}_{m, :}, J_{m}| \bm{X}) f(\bm{y}_{m} | \bm{z}_{m, :}, \bm{X}, \bm{k}_m)  f(\bm{t}_{m} | \bm{z}_{m, :}, \bm{X}, \bm{e}, \bm{k}_m) ,
\end{array}
\end{equation}
where the summation of latent call capture history $\bm{z}_{m,:}$ at detector $m$ is constrained by the detection number $J_{m}$. We use a Metropolis-Hastings (MH) algorithm to sample $\bm{z}_{m, :}$ and $\bm{k}_m$ together. 

The proposal distribution $q(\bm{z}_{m, :}, \bm{k}_m)$ is set to be a discrete Uniform distribution over all possible combinations of latent capture history $\bm{z}_{m,:}$, and the detection order $\bm{k}_m$ is naturally determined by $\bm{z}_{m,:}$, $\bm{X}$, and $\bm{e}$. 
Since the proposal distribution is an independent sampler with a discrete Uniform distribution, we have $q(\bm{z}_{m,:}, \bm{k}_m) = q(\bm{z}^{*}_{m,:}, \bm{k}^{*}_{m})$, where the random variables with superscript $^*$ mean the newly sampled value in the MCMC iteration. Thus we accept the proposal with the probability
\begin{equation}
     min \left\{ 1, \quad \frac{f(\bm{z}^{*}_{m, :}, \bm{k}^{*}_{m} |  \bm{X}, J_{m}, \bm{y}_{m}, \bm{t}_{m}, \bm{e})}{f(\bm{z}_{m, :}, \bm{k}_m |  \bm{X}, J_{m}, \bm{y}_{m}, \bm{t}_{m}, \bm{e})}   \right\} . 
\end{equation}
\vspace{1em}

\subsection*{Sampling Detail of Latent Location $\bm{X}$}

The sampling posterior for the latent location is 
\begin{equation}
   f(\bm{X} | \bm{Z}, \bm{Y}, \bm{T}, \bm{e}, \bm{K}) \propto f(\bm{X}) f(\bm{Z} | \bm{X}) f(\tilde{\bm{Y}} | \bm{Z}, \bm{X}) f(\tilde{\bm{T}} | \bm{Z}, \bm{X}, \bm{e}) .
\end{equation}
We omit $\bm{J}$ from sampling posterior since the latent location is conditionally independent of the detection number when the latent capture history exists. 
Since the prior for each call latent location is assumed to be independent, we have the sampling posterior for each call location as 
\begin{equation}
f(\bm{x}_{n} | \bm{z}_{:,n}, \tilde{\bm{y}}_{:, n}, \tilde{\bm{t}}_{:,n}, e_{n}) \propto f(\bm{x}_n) f(\bm{z}_{:,n} | \bm{x}_{n}) f(\tilde{\bm{y}}_{:, n} |  \bm{z}_{:, n}, \bm{x}_n) f(\tilde{\bm{t}}_{:,n} |  \bm{z}_{:, n}, \bm{x}_n, e_{n}) .
\end{equation}
We also use an MH algorithm to sample the location $\bm{x}_{n}$, and the $\bm{K}$ is omitted since the $\tilde{\bm{Y}}$ and $\tilde{\bm{T}}$ have all the information in detection order.

To have a better mixing on the Markov chain, we apply proposal distribution as a mixture of the Uniform distribution and the Gaussian distribution:
\begin{equation}
   \bm{x}_{n}^{*} \sim w N(\bm{x}_{n}, \sigma_g) + (1 - w)U(A) ,
\end{equation}
where $w$ is the mixture weight and $\sigma_g$ is the scale parameter. These hyper-parameters are set based on the previous work on \citet{10.1093/biomtc/ujad019}. Since each component of the mixture proposal is symmetric, i.e., $q(\bm{x}_{n}^{*} | \bm{x}_{n}) = q(\bm{x}_{n} | \bm{x}_{n}^{*})$, we have a Metropolis sampler with acceptance probability as 
\begin{equation}
   min \left\{1, \quad \frac{f(\bm{x}^{*}_{n} | \bm{z}_{:,n}, \tilde{\bm{y}}_{:, n}, \tilde{\bm{t}}_{:,n}, e_{n})}{f(\bm{x}_{n} | \bm{z}_{:,n}, \tilde{\bm{y}}_{:, n}, \tilde{\bm{t}}_{:,n}, e_{n})}   \right\} .
\end{equation}
\vspace{1em}

\subsection*{Sampling Detail of Call Emission Time $\bm{e}$}
The sampling posterior for the latent call emission time is
\begin{equation}
        f(\bm{e} |\bm{X}, \bm{Z}, \bm{T}, \bm{K}) \propto  f(\bm{e})  f(\tilde{\bm{T}} | \bm{Z}, \bm{X}, \bm{e}) .
\end{equation}
Since the prior for the call emission time is assumed to be independently Uniform distributed, and the call detection order $\bm{K}$'s information is included in $\tilde{\bm{T}}$ matrix, we have the posterior for individual call emission time as
\begin{equation}
    f(e_{n}| \bm{x}_n, \bm{z}_{:,n}, \tilde{\bm{t}}_{:,n}) \propto f(e_{n}) f(\tilde{\bm{t}}_{:,n} |  \bm{z}_{:, n}, \bm{x}_n, e_{n})
\end{equation}
Since the prior $f(e_{n})$ is uniformly distributed, and the $f(\tilde{\bm{t}}_{:,n} |  \bm{z}_{:, n}, \bm{x}_n, e_{n})$ is a product of independent Normal distribution in each detector, we have an analytical posterior as a truncated Normal distribution:
\begin{equation}
    e_{n} \sim \psi(\mu_{e_{n}}, \sigma_{e_{n}} , e_{l}, e_r) ,
\end{equation}
where we can directly sample $e_{n}$, and
\begin{equation}
\begin{array}{ll}%
    \mu_{e_{n}} &= \frac{1}{\sum_{m =1}^{M} z_{m,n}} \sum_{m: z_{m, n} = 1} \tilde{t}_{m,n} - d_{m}(\bm{x}_n) / v \\
    \sigma_{e_{n}} & = \frac{\sigma_{t}}{\sqrt{\sum_{m =1}^{M} z_{m,n}}} , 
\end{array}
\end{equation}
where $v$ is the sound speed and $e_l, e_r$ is the left and right boundary on possible call emission time.

\vspace{3em}

\section*{Web Appendix D - Bootstrap Details}
During the bootstrap, the call location dependency is simulated using the pdf of the separately estimated target animal call rate $\hat{f}(\mu_c)$. The estimate of the animal density is given by $\hat{D}_a = \hat{D}_c / \hat{\mu}_c$.
The details of the bootstrap algorithm are listed below (The simulated data or parameters estimated from simulated data are denoted with the superscript $^*$):

\begin{enumerate}
\item Simulate animal location as a realization of a homogeneous Poisson process with intensity $\hat{D}_{a}$.

\item Generate $\bm{X}^{*}$ by repeating each location from step (1) $\mu_c$ times, where $\mu_c$ is the call rate drawn from pdf $\hat{f}(\mu_c)$ and the call rate distribution $\hat{f}(\mu_c)$ is estimated from an independent dataset. 

\item Obtain $\bm{Z}^*$ by simulating from the estimate of $f(z_{m, n} | \bm{x}_{n}^{*})$ with probability function (\ref{equ:det_auto}). 

\item Obtain $\tilde{\bm{Y}}^{*}$ by simulating from the estimate of $f(\tilde{y}_{m, n}|z_{m, n}^{*} = 1, \bm{x}_{n}^{*})$ with Equation (\ref{equ:ss_auto}) for all observations.

\item Obtain $\bm{e}_{n}^{*}$ by sampling from a Uniform distribution in survey duration and $\tilde{\bm{T}}^*$ by simulating from the estimate of $f(\tilde{t}_{m,n}|z_{m,n}^{*} = 1, \bm{x}_{n}^{*}, e_{n}^{*})$ with Equation (\ref{equ:toa_pdf}) for all observations.

\item Obtain $\bm{J}^{*}, \bm{Y}^{*}, \bm{T}^{*}$ by removing all matching information from $\bm{Z}^{*}, \tilde{\bm{Y}}^{*}, \tilde{\bm{T}}^{*}$ on each detector.

\item Estimate $\hat{\bm{\gamma}}^{*}, \hat{\bm{\phi}}^{*}, \hat{D}^*_c$ from $\bm{J}^*$, $\bm{Y}^{*}$, and $\bm{T}^{*}$ using proposed MCEM algorithm.

\item Calculate $\hat{D}^*_a$ with $\hat{D}^{*}_a = \hat{D}^{*}_c / \hat{\mu}_c$.

\item Repeat the above steps $B$ times and save the parameter estimates from each iteration.

\end{enumerate}

\vspace{3em}

\section*{Web Appendix E - Software Implementation}
The proposed model is implemented using R with package \textit{ascr}~\citep{ascrPackage}, and Julia language with package \textit{Optim} \citep{Optim.jl-2018}, where \textit{ascr} is used for generating simulated data and acquiring frog survey dataset, and \textit{Optim} is used for optimization of the ELBO function in M-step.

\label{lastpage}

\end{document}